# Planar waveguide nanolaser configured by dye-doped hybrid nanofilm on substrate

E.A. Tikhonov[a*], V.P. Yashchuk[b], G.M. Telbiz[c]

[a]Department of Coherent&Quantum Optics, Institute of Physics, NA Sci. of Ukraine,
46 Nauki av., Kyiv 03028, corresponding author e-mail: etikhon@gmail.com

[b]Department of Physics, T. Shevchenko National University, Kyiv 01601, Ukraine

[c]Department of Physical-Chemistry, Institute of Physical Chemistry, NA Sci. of Ukraine, Kyiv 03039

**Abstract**

Dye-doped hybrid silicate/titanium nanofilms on the glass substrate in the structure of asymmetrical waveguides were studied as the laser system. The spatial and spectral features of laser oscillation of genuine and hollow waveguides were determined. Their lasing thresholds of a two-order difference of pumping power were identified. The calculated cutoff thickness for waveguides of ≈100 nm did not exceed their actual thickness of 300nm. The overall pattern of stimulated emission included two concurrent processes: one-mode waveguide lasing and the lateral beam emission of small divergence emanated through the waveguide wall as leaky modes. The comparison of the open angle of lateral beams and grazing angles of the waveguide lasing mode provided insight into the effect as the leaky mode emission converted in the Lummer-Gehrcke interference.

# 1. Introduction

The inclusion of organic molecules in the solid inorganic matrix at high concentration but the minimal interaction to keep their luminescent efficiency is the weighty reason for the promotion of similar research and development. The high concentration of the luminescent centers is a useful feature of laser materials because actually, dopant determines the specific amplification or nonlinear response (dye stream lasers, dyed liquid crystal laser, harmonic generation, and others). The solution of this task is possible with the structured inorganic matrix. The presence of matrix cells (pores) permits to place organic molecules separately. When the matrix pore concentration is comparable to guest molecules, the interaction among them gets attenuated.

For the first time, the regular mesopores (i.e. filled by intermediate matters) in silicate glasses were performed by sol-gel technology by authors [1, 2, 3]. The first porous matrixes created by acid-alkaline etching of soda-borate-silicate glasses [4]. The dye deposition into the matrix pores performed by the sorption from the solution. In this approach "individual cells" for separated molecules were absent. Similar dye-doped solid matrix with the increased thermo-optical stability has been gained in this way and used successfully. Frequency tunable lasers of the similar dye-doped meso-porous matrixes have been realized in a few works early [5].

Another room temperature sol-gel technology of chemical reactions for the production of hybrid laser material with random pores has been embodied [6]. The output product here was called "ormosil" and was gained in the usual 3-dimensional form. Significant research has been performed with lasing on the dye-doped ormosil matrixes at typically low concentrations precluding the negative consequences of the dye dimerization [7].

In our work, the structured dye-doped material has been produced with the modified sol-gel technology. The technology relies on the ability of amphiphilic molecules to self-assembling in modular structures – micelles. In a time of dehydration stage self-organizing of dyed micelles in spatially ordered ensembles of various symmetry takes place. The concurrently with the mentioned process, the relevant structuring of silicon/titanium matrix transpires. With this technology, the dye-doped hybrid material in the form of extended films of the nanoscale thickness is produced. Herewith given films demonstrate the higher luminescent quantum yield at dye concentrations more higher than the usual impurity solution and solids do [8,9]. Changing concentration relation between organic-nonorganic parts of material recovers the way to control of index refraction (IR), the thickness of films and some other properties. These properties of the dye-doped hybrid nanofilms were a reason of given application them as nanoscale laser media.

The miniaturization of lasers to nanoscale dimensions is the subject of the heightened research in last time. It provides small mode volumes and strongly enhanced fields as a prerequisite of nonlinear interaction followed by appropriate nonlinear phenomena. 1-dimensional semiconductor nanolasers were investigated in works [10-12]. Hybrid plasmonic zinc oxide nanowire laser near the surface plasmon frequency was considered in publication [13]. The 3-dimensional fully dielectric laser has been proposed and theoretically justified in work [14]. As cavity of this designed nanolaser is proposed to take a dielectric silicon sphere covered gain layer with diameter 100-200nm.

In real work, dye-doped nanoscale films on a substrate it is proposed to be used as part of 1-dimensional laser system based on one-mode asymmetrical waveguide as the most appropriate cavity. Consequently, the given work is targeted to the creation and overall investigation of 1-dimensional nanolasers based on the dye-doped hybrid organic-nonorganic nanofilms.

## 2. Characterization of dye-doped nanofilms

The mesoporous hybrid $SiO_2/TiO_2$-films were fabricated by the sol-gel technology using surfactant nonionic three-bloc copolymer Pluronic P123 (from Sigma-Aldrich). A sol-solution was prepared using chemically pure tetra ethoxy silicate ($Si(OC_2H_5)_4$) or titanium ($Ti(C_2H_5)_4$), ethanol with distilled water and 35% HCl. After mixture hydrolysis and sol origination, the made before gel of R6G and surfactant Pluronic P123 was mixed at a first critical concentration. This step resulted in the self-assembling of micelles in the random spatial distribution. A ratio between the gel and intermixture concentrations in gram-molecule units was varied from 1:32 to 32:1. The basic regular composition of the material was agreed to next component ratio:
teoSi(Ti):$C_2H_5$OH:HCl:$H_2$O=1:20:0.5:8. The final intermixture was executable to film formation after aging at the continuous stirring during (40÷60)min. A film deposition on the glass substrate was carried out at the gyration and drive velocities of 2040 round/min and (0.1-20) cm/min, respectively. The deposited silicate (titanium) films are dried at the regular temperature/damp and until their thicknesses reached values of 100-300nm. Due to drying concentration of dyed micelles in the film increased to the second critical value. As result, the spatial ordering of dyed micelles takes place simultaneously with the $SiO_2$ ($TiO_2$) grid structure formation. The grid walls of (3÷5) nm are built around the spatially ordered micelle structure. Micelle ordering manifests in the plane and across the nanofilm. At conservation of the organic micelles in the gained film, the similar structure is referred the mesoporous. Heating of similar samples up to 500°C provides the evaporation of the organic micelles and, respectively, the formation of the true porous films.

The formation of the ordered structure confirmed by the small-angle x-ray diffraction technique. The results confirm the formation of the ordered structure and its invariance at the supplementation of non-amphiphilic laser dyes. Infrared spectroscopy has specified almost full elimination of water from the films after drying. The film thickness was in the range of (100÷300)nm according to the measurement of the atomic force microscopy. The next plain estimation provides the parameters of the shaped spatial structures. Diffraction maximum on x-ray radiation ($\lambda_{CuK\alpha}$=0.15405nm) satisfies to Bragg condition $m\lambda=2\Lambda\sin(\Theta_m)$ at the discrete diffraction angle. It permitted to determine the spatial periods of the grating $\Lambda_1$=10,204 nm and $\Lambda_2$=10,155nm for m=1 and 2 respectively.

Herewith the density of micelle packing is about $\approx 10^{18}$ cm$^{-3}$ for the cubic symmetry if the spatial period and orderliness of the structured $SiO_2$ films correspond to the spherical micelles. The micelle packing density is about one order higher for hexagonal packing. At the measured optical density **cεT**=1,5 (dye R6G, molar extinction $\varepsilon_{max}$=1,2*10$^3$, film thickness T=200nm) in the maximum of absorption band, the dye concentration is 0,6m/L=3,6*10$^{20}$cm$^{-3}$. The comparison of micelle and dye concentrations shows about ten-fold excess of dyes above pores. But the shapes of the dye absorption bands did not show common signs of dimerization.

The measurements of the index refraction (IR) for dye-doped $SiO_2$ and $TiO_2$ were conducted with the modified Brewster [15] and traditional ellipsometry methods, respectively. Both optical structures satisfy to next configuration of layer interleaving "air-film-glass substrate" which is relevant to the asymmetrical planar waveguides /16/. Brewster measurements of IR of nanofilms were performed at two wavelengths: 532nm and 660nm. A smaller wavelength 532nm overlaps with the absorption maximum of R6G in $SiO_2$ and a larger one (660nm) fits the fluorescence band. The real n and imaginary ϰ parts of the complex IR become commensurable when the concentration of R6G in the nanofilm reaches relatively high level. Then measurements of the complex IR by Brewster method become ambiguous compared to the common case: **n>>ϰ=λα/2π**. The ambiguity can be removed by independent measurement of the relevant absorption coefficient (α). In the case when the strong absorption attenuates of the second reflected beam at 532nm from the substrate the light reflection curve was observed without interference accompaniment (Fig.1a). The measurement outside of absorption band at 660nm shows the genuine Brewster angle for the nanofilm with Pluronics123 + dye (Fig.1b). The two-beam interference accompanies the relevant curve due to the presence of the two beams in the reflection, but it did not affect on accuracy.

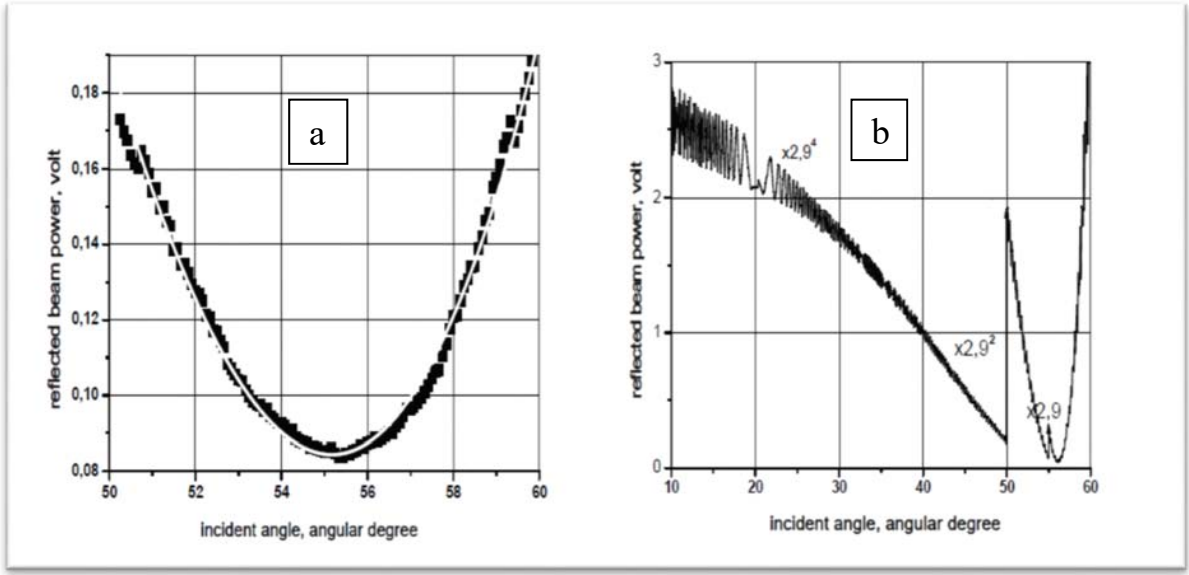

Fig.1 Determination of IR using the Brewster minimum of reflection power on the wave TH-polarisation for the dye-doped mesoporous $SiO_2$ film: a - inside absorption band at 532nm and n=1,438; b – inside the fluorescent band at 600nm and n=1,491.

The Brewster angular minimum in the absorption band depends from the complex IR according to the relation **n=1/cos($\phi_{min}$)√(1+$\varkappa^2$)** and **$\varkappa=\lambda\alpha/2\pi$** /15/. The waveguiding of the planar structure is satisfied when the condition **$n_f$ >$n_{c,s}$** becomes valid. Here **$n_{c,s}$** means IR for substrates and cover, **$n_f$** is IR of the relevant mesoporous film. So fabrication of the mesoporous film with application of $TiO_2$ (IR=2,62) was imperative. IR of the hybrid film reduces to value ≈ (1,58-1,68) because the organic part of the film remained in kind but can fluctuate partly.

There is the definite interest to the dependence of absorption/luminescence spectrum at various dye concentrations in films. The study has shown until the dye concentration stays below $10^{-4}$M no deviations from the typical monomolecular spectrum of R6G in water take place. The evolution of the fluorescence spectrum with the concentration change for hybrid films $SiO_2$ and $TiO_2$ are presented on Fig.2. [8,9]. The fluorescence efficiency R6G in the nanofilms conserves until dimerization will starts.

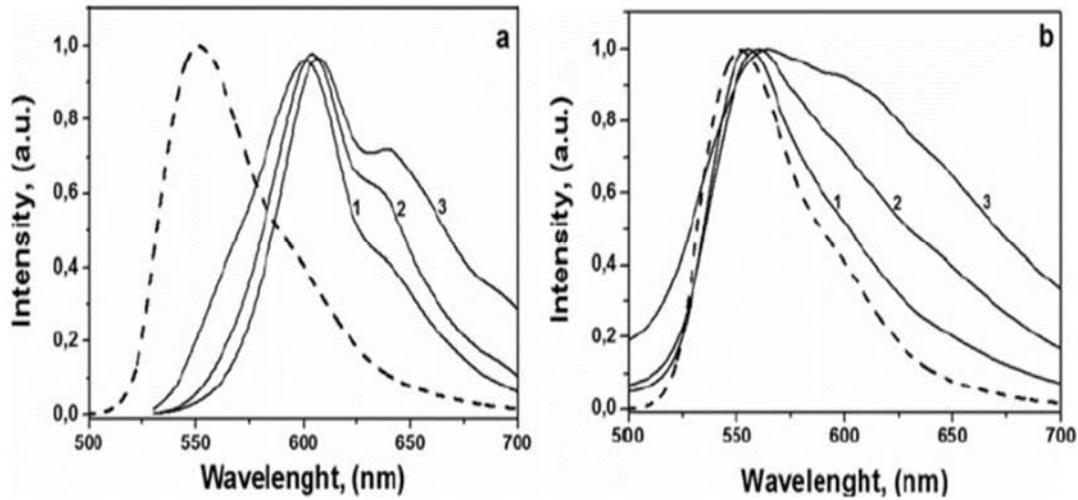

Fig.2 The normalized fluorescence spectra of R6G in Pluronic123/SiO$_2$(a) and Pluronic123/TiO$_2$ (b) hybrid sol-gel films at concentrations: (1) 0,41mg/ml; (2) 3,6 mg/ml; (3) 10 mg/ml. The dashed line is fluorescence spectra of dyed ethanol solution (c = 10$^{-5}$ M).

### 3. Experimental laser procedures, results, discussion

The considered dye-doped nanofilm on a glass substrate has shaped the planar waveguide. This waveguide configuration has the structure of the asymmetrical waveguide due to IR difference of the substrate and the cover. The main features of the planar dielectric waveguides presented in [16] and other cited there works. To describe the waveguide behavior in our case the ray-optical approach turns to be enough. According to the "zig-zag" path model, the wave propagation has two groups of orthogonal modes: TE and TH types. If the light incident angle inside of the core film satisfies $\Theta_I < Q_{TIR}$ (TIR is a total internal reflection), the light reflection with R<1 takes place without phase shift but with jump $\pm\pi$. When the incident angle exceeds the angle $Q_{TIR}$ reflection reaches 100% and includes the variable phase shift. The phase shift grows when $\Theta_I$ exceeds $Q_{TIR}$ and reaches $\pi/2$ at $\Theta_I = \pi/2$. Last features describe the classic Fresnel theory. The particular features of the light propagation in asymmetric waveguide depend on the relation between of grazing angle and two different TIR angles. At $\Theta_I < Q_{TIR1}, Q_{TIR2}$ light leaks from the waveguide and makes leaky mode propagation. The same is valid also when $\Theta_I \lesssim Q_{TIR2}$, but light emanates just from the substrate. So necessary condition of existence of guided modes is the next inequality: $\Theta_I > Q_{TIR1}$,

Q$_{TIR2}$. Sufficient condition of guided (or leaky) mode existence appears when the sum of all phase shifts along the "zig-zag" trajectory gets equal to multiple number 2π. The self-consistency condition (or transverse resonance) takes the following form:

$$2kn_f T \cos\Theta_I - 2\vartheta_s - 2\vartheta_c = 2m\pi \quad (1)$$

Only a discrete set of TIR angles satisfies to a self-consistent surviving of guided and leaky modes. Here k is a wave number, 2ϑ$_s$ and 2ϑ$_c$ are phase shifts for waves in the substrate and cover interfaces respectively. They are taken with minus sign because of the time delay at the reflection, m=0,1,2,3.. are numbers of the modes. The phase shift in the first member of (1) is inversely proportional to the grazing angle Θ$_I$.

The arbitrary small thickness T is permitted by this equation (includes T=0) for the symmetrical waveguide on main guided mode (m=0). This implies there is no cut-off frequency for the fundamental mode. But multimode operation (m>0) according to (1) is allowed only for аштфд waveguide thickness T≠0. For the asymmetrical waveguide equation (1) shows that waveguide thickness has to be T≠0 even for one-mode operation. The decrease of core thickness is followed by a decrease of a grazing angle Θ$_I$. As Θ$_I$ approximates larger angle Q$_{TIR2}$ the respected phase shift reaches zero and stops to change. From this region of grazing angles, the guided mode operation will change by the leaky mode one. This means that cutoff thickness on the main mode of the asymmetrical waveguide can be found from relation (1) at condition 2ϑ$_s$=0:

$$T_c = \vartheta_c / kn_f \cos\theta_I \quad (2)$$

The phase shift of TE- polarized wave at an incident angle above TIR2 angle, according to Fresnel theory is determined:

$$\vartheta_c = \sqrt{(n_f^2 \sin^2\theta_I - n_c^2)}/n_f \cos\theta_I \quad (3)$$

Now the determination of the cutoff thickness of asymmetrical waveguide T$_c$ is getting plain because the grazing angle is equal to the bigger angle Q$_{TIR2}$. Taking into account that phase shift ϑ$_s$≈0 and substitution (2 and 3) in (1) :

$$T_c = \lambda \arctan \sqrt{n_f^2 (\sin^2 \theta_{TIR2} - \sin^2 \theta_{TIR1})} / 2\pi n_f^2 \cos^2 \theta_{TIR2} \qquad (4)$$

When the difference between angular values TIR2 and TIR1 approximates zero, the cutoff thickness of waveguide core $T_c$ approaches to zero also: it means allegedly the asymmetrical waveguide transforms to the symmetrical one. The ratio of $T_c/\lambda$ is called wavelength or frequency cutoff and it shows the smallest "part" of the wavelength which emission can be transferred with the given asymmetrical waveguide.

Our genuine asymmetrical waveguide on the dye-doped hybrid $TiO_2$ film with IR $n_f \approx 1,68$ on the glass substrate ($n_s$=1,51) has the angles of $Q_{TIR2}$=64$^0$ and $Q_{TIR1}$=36,53$^0$ on the interfaces of the film-substrate and the film-air respectively. The evaluation of $T_c$ by the expression (3) at $n_f$=1,68 provides the cutoff thickness $T_c$=131nm. The reduction of film IR up to 1,58 gives an increase of TIR2 up to value $Q_{TIR2}$= 72,88$^0$. The cutoff thickness drops to $T_c$= 80,92nm in this connection.

The hollow asymmetrical waveguide is accomplished with the dye-doped hybrid $SiO_2$ film. This film has wavelength dependent IR=1,49 (Fig.1) that is less IR of the glass substrate $n_s$=1,51. The given waveguide is characterized by the higher radiation losses and all guided modes are leaky. In spite of that under amplification existence lasing process can be excited on the leaky guided 17,18].

The equation (1) can apply for the description of the discrete angular distribution of leaky modes of the hollow waveguide also. In that case, the light transfer process has two parts on leaky modes: the first part propagates inside of waveguide core and the second part of the power is emitted outside through the lateral wall. The phase jump on $\pi$ at the light reflection from more dense substrate media has to be taken into account. The grazing angles of leaky modes can be found from the presented equation (1):

$$\Theta_m = \arccos((2m+1)\lambda / 4n_f T) \qquad (5)$$

where m=0,±1,±2 …. and the grazing angles of the fundamental mode at m=0 $Q_0$ are 59,8$^0$ and 70,4$^0$ for two core film thicknesses T of 200nm and 300nm, respectively, $n_f$=1,491, $\lambda$=600nm. The highest modes m>0 are forbade because the condition of the transverse resonance is not doable for the given film thickness.

The spatial and spectral features of stimulated emission of the planar asymmetrical waveguide with different IR of the core nanofilm are considered below. There were two types of the asymmetric waveguides with the core dye-doped R6G at the approximately equal concentrations. As noted above, $SiO_2$ hybrid core film forms a hollow waveguide and the $TiO_2$ hybrid core film forms the genuine waveguide. The transverse laser pumping (532nm, ≈30ns) at beam cross-section 18x0.1mm$^2$ injected in the waveguide through its lateral wall. The pump intensity might change stepwise by attenuators within the confines 0.05 – 0.35 MW/mm$^2$. The stimulated emission of high diffraction divergent has been excited inside the waveguide dye-doped core film. Two symmetrically located lateral beams of the much lesser divergence emerged under the pumping also. Named beams intercepted by the optical system in turn and directed alternately to the spectrograph for analysis. The divergence of the main waveguide radiation appeared naturally in the plane orthogonal to the waveguide plane. Herewith divergence in the plane of the waveguide remained lower and determined by the shape/size of the excited film region. Next measurements were performed for the comparison of the stimulated emission (threshold, angular and spectral characteristics) of the genuine and hollow waveguides under variable pumping.

It is well known that luminescence of transparent films owing TIR becomes much stronger around all side ends in compare to wide plane surfaces. With the enhanced excitation of the similar planar "waveguide" luminescent power will increase much faster than for the systems without TIR when the light can propagate free in 4π steradian. Most likely that the conditions and thresholds of laser oscillation are more optimal in the genuine waveguide than in the hollow one. Herewith the laser oscillation requires the conditions for survival of the minimal number of the available modes of spontaneous emission. These conditions are known as the phase and amplitude lasing. Equation (1) displays the phase conditions that highlight the lowest loss of guided modes in the waveguide. The radiation losses of the waveguide modes must compensate the amplification. The spectrally dependent feedback may appear also through counter-propagated ASE-beams (amplified spontaneous emission or superluminescence). No wavelength selective elements are nesessary if the coupling between the counter running waves owing to interference and the gain grating recording takes place.

All observed types of stimulated emission in the asymmetrical hollow waveguides presents schematically Fig. 3. The excited region of the dye-doped waveguide core film has emitted the specular symmetrical spatial pattern of the stimulated radiation. The central lobe of the pattern is characterized by a relevant diffraction-limited divergence in the plane orthogonal to the waveguide plane $\lambda/T \approx 2$rad at T= 300nm and $\lambda \approx 580$nm. The angular divergence of the central lobe in the orthogonal waveguide plane is less than $1^0$. This divergence depends on just the pumping spot geometry. The total patterns of the spatial beam distributions for hollow and the genuine waveguides differs only slightly despite the different properties of leaky and guided modes.

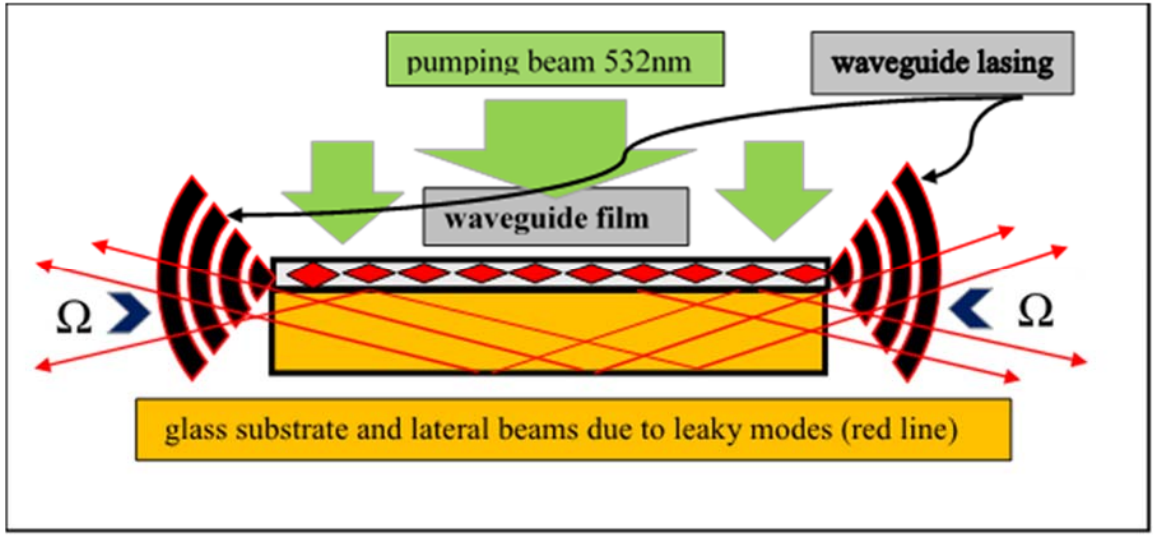

Fig.3. The sketch of lasing waveguide structure and overall beam pattern of lasing emission

The main lasing beam demonstrates the sharp spectral narrowing from the initial broad luminescence band of R6G $\approx 550$nm to the lasing linewidth $\approx 4$nm when pumping reaches the threshold levels (Fig.4a,b). It is well known lasing process differs from ASE by sharp spectral narrowing. The same dramatic narrowing with pumping manifest stimulated emission for the genuine and hollow waveguides. However, lasing thresholds (in points of spectral narrowing) for the genuine and the hollow waveguide were differed by $\approx 1.5$ order. The reason should be connected low radiation losses in the genuine waveguide in comparing the same losses in hollow one. Fig.4a illustrates also the behavior of the spectral lines when the pumping is over the threshold value. One can be seen that the spectral bandwidth remains approximately the same.

    The stimulated emission of lateral beams resolutely differs from waveguide lasing mode (m=0) in the respect to spectral widths and angular divergences (Fig.3). These lateral beams in both types of waveguides have a smaller angular divergence ($\approx 1^0$) in the waveguide plane and

bigger divergence in the orthogonal plane as compared to the main mode emission. The low angular divergence of the lateral beams suggests immediately another output emission spot than the face of the waveguide core. As an output spots of lateral beams, it is suggested to consider the side wall of the waveguide. But this way is difficult to explain the origin of the lateral beams outside the genuine waveguides.

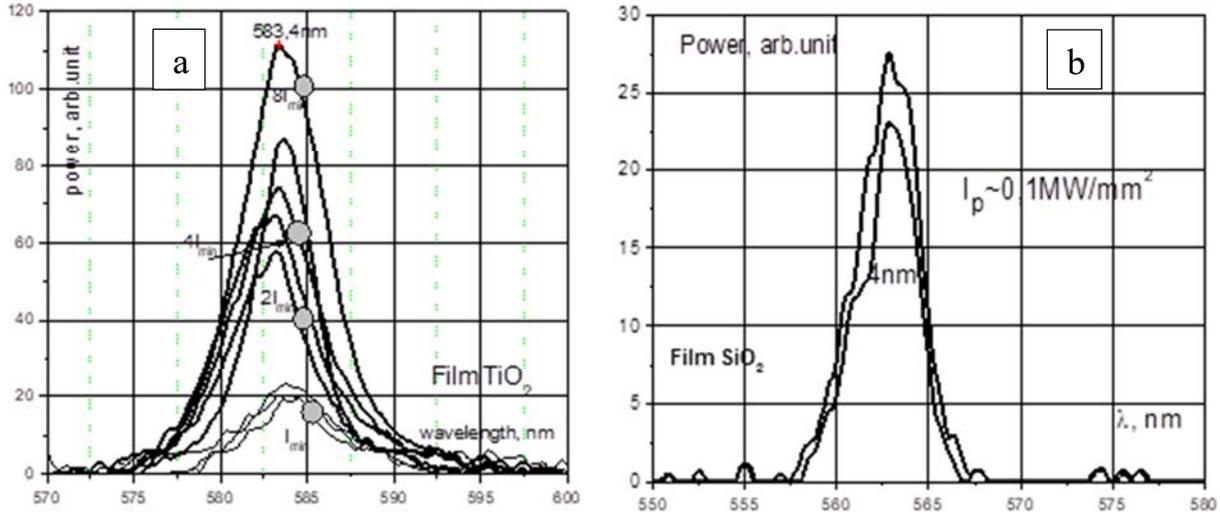

Fig.4. a) The power of the spectral distribution for genuine waveguide made of dye-doped $TiO_2$ film at the various pumping for lasing stimulated emission in the central lobe;
b) The same spectral distribution for the hollow waveguide made of the dye-doped $SiO_2$ film ht).

The lateral beams for both waveguides appeared with opening angles about $52^0$ in the same plane where all beams overlapped by each other. The conclusions about the beam overlapping were pulled after the comparison of the spectral distributions of the single and overlapped beams. The emission spectra outside the beams have manifested the known luminescence spectrum of R6G (see Fig.2.). The emission of lateral beams under angle about $\pm 26^0$ has shown the small smooth narrowing of the initial broad R6G spectrum - like to evolution of ASE spectra versus pumping. The spacial overlapping of the waveguide lasing emission and lateral beam emission (see fig.3.) is developed in their mutual spectrum: they have the different spectral bandwidth.

The small angular divergence of the lateral beams results from their emanation through the lateral waveguide walls without diffraction angular broadening. They should be considered as waveguide leaky modes. The next summation of some successive coherent beams along the excited waveguide region transforms to Lummer-Gehrke multibeam interference pattern [17,18]. The pattern is unwrapping in the plane parallel to the waveguide plane with a divergence about $20^0$ (Fig.3.). Fewer questions have arisen for comprehension the lateral beams in the case of the hollow waveguide. But for the genuine waveguide, it is necessary to suggest at least the partial TIR violation! To be consistent with the phase equation (1) it may be supposed the small deflection from the angle of TIR for the grazing angle of the lasing mode. Some support in favor this thought shows the distinction of the spectral widths of main lasing line and ASE in lateral beams.

Furthermore, consider the correlation between the open angle for the lateral beams $\Omega \approx 52^0$ ( $\pm 26^0$ to the axis of the waveguide core assigned the pumping beam spot, Fig.3.) and the grazing angle for the waveguide lasing mode. The measured open angle $\Omega$ leads to finding the grazing mode angle using Snell law. At the same time, the grazing mode angle can be calculated from equation (1) and known values T and IR of the dye-doped film. Equality of those angular values should confirm the joint origin of lasing mode and lateral beam emission.

The next results have been gained on the above-mentioned notes. Taking into consideration the initial incident angles $\pm 26^0$ and its succeeding refraction on transits across substrate up to film core it was found the angular values $\pm 73^0$. The grazing angle for the main mode of the hollow waveguide at the film thickness T=300 nm, $n_f$=1,491 and $\lambda$=600nm equals $\pm 70,4^0$. The deficiency of accurate data about the thickness used $SiO_2$ - nanofilm made no sense to perform more accurate comparing.

The explanation of the lateral beam origin as leaky mode at the same open angle $\pm 26^0$ in the case of the genuine waveguide was more complicated in comparison to the hollow one. Small violation of TIR condition at accidental overlapping angles $Q_{TIR2}$ and grazing mode proposed above as some solution. The determination of the incident angle inside film core from the lateral beam directory again leads to the value $\pm 73^0$. Now it was necessary to settle the TIR2 angle and the grazing mode angle of the genuine waveguide based on $TiO_2$ nanofilm. The angular consistency was achieved for the film of the same thickness T=300nm and IR=1,58 (instead of expected IR=1,68) that resulted in the grazing angle $\pm 70,4^0$ for the main mode m=0 and $Q_{TIR2}$= $72,88^0$. Deficiency of more accurate value IR used $TiO_2$-nanofilm made again no sense for the more accurate comparing of experimental and calculated magnitudes now.

## 4. Summary

The stimulated emission based on the dye-doped nanofilm of (200±300) nm thickness in the structure of the planar asymmetrical waveguide is discovered and studied for the first time. The waveguide nanofilms formed on the glass substrate by the specific chemical deposition. The stimulated emission reveals the complicated structure that depends on either genuine or leaky waveguide operates. The evaluation of the wave cutoff of asymmetric genuine waveguides at thickness T=300nm has supported the possibility of the guided mode operation for the thickness above $T_c$ =130 (81) nm. The studied core nanofilms belong to the class of hybrid organic/nonorganic materials (structured ormosil) with the support of ordering due to the organic micelles inside the inorganic matrix. The x-ray diffraction confirms the ordering of the inorganic matrix. The micellar structure of the organic part of the material sets a quasi-structured skeleton with the embedded dye in the monomolecular state. Due to dyes luminescence efficiency preserves at higher concentrations than for the usual system. The hybrid $SiO_2$ film forms the hollow waveguide on the glass substrate and operates in the leaky modes. The hybrid $TiO_2$ film works as a genuine waveguide on the good quality guided modes. The calculation has shown that both types of waveguides operate on the main single-mode. The measurements confirm that lasing for the genuine waveguide has the laser threshold about 2 orders lower than for the hollow one. The integral of the spatial structure of radiation includes the lateral beams of ASE with relatively low angular divergence. This radiation has referred to the leaky Lummer-Gehrke interference modes. A broader spectral bandwidth and low divergence of this ASE are due to both beams located outside of a waveguide lasing core and emanated from the interface film-substrate. It is quite evident for hollow waveguide but is not evident for the genuine waveguide. The estimations have shown that leaky mode can appear if waveguide film parameters make the grazing angle for the guided mode nearly equal to the bigger angle of TIR. The future study would be done for the best clarification of this observation.

Technological/physical difficulties and perspectives for the next development of the dyed structured ormosil films and their application in nano-optics are on the agenda. One of the promising conclusions the given project is applicability the waveguide structures for testing and development new laser material and nanolasers.

### Acknowledgements

This research has been supported by the National Academy of Sciences of Ukraine (NASU) in the frame of the project # 0106U002454 and fulfilled by the Institute of Physics NAS Ukraine in the collaboration with the Institute of Physical Chemistry and Department of Physics at T. Shevchenko National University.

**References**

1) Kresge C., Leonowicz M., Roth W., Vartuli J., Beck J., 1992, *Nature*, **359**, 710-715.

2) Beck J., Varuli J., Roth W., et al.,1992, *J. Am. Chem. Soc.* **114**, 1083-88.

3) Zhao D., Huo Q., Feng J., Bradley F., Galen D., 1998, *J. Am. Chem. Soc.* **120**, 6024-2030.

4) Zhdanov S. *About a structure of a glass according to an examination of the structure of porous glasses and films,* 1955, Publishing house Acad. of Sci., USSR.

5) Altshuler G., Dul'neva E., Yerofeev A., 1985, *Journal Tech. Physics, **55(8)***, 1622-1630.

6) Proposito P., Casalboni M*., 2003, Handbook of Organic-Inorganic Hybrid Materials and Nanocomposites,* I, XXX, Amer. Sci. Publisher.

7) Dunn B., Mackenzie J., Zink J., Stafsudd O., 1990*, Proc. SPIE*, Sol-Gel Optics, **174,**1328-1335, (doi:10.1117/12.22557)

8) Tikhonov E., Telbiz G., 2011, *MCLC*, **535**, 82-90, (doi:10.1080/15421406.2011.537950)

9) Telbiz G., Leonenko E., Dvoynenko M., 2017, *MCLC*, **642:1**, 74-80, (doi:10.1080/ 15421406.2016.1255052)

10) Yan R., Gargas, D., Yang, P. 2009, *Nature Photonics* **3**, 569–576.

11) Ma Y., Guo, X., Wu, X., Dai, L., Tong, L. 2013, *Advances in Optics and Photonics,* **5**, 216.

12) Saxena D., Mokkapati S., Parkinson P. and others, 2013, *Nature Photonics* **7,** 963–968

13) Themistoklis P., Sidiropoulos H., Röder R., at al 2014, *Nature Physics,* **10,** 870–876

14) Kewes G., Rodríguez-Oliveros R., Höfner K. at al **2014**, *arXiv:1412.4549v2*

15) 10) Tikhonov E., Lyamets A., **2015,** *arXiv.org:1509.09191*

16) Kogelnic H., *Theory of Dielectric Waveguides,* 1975, Integrated Optics, **7,** 13-81, Springer Verlag, Berlin Heidelberg, Academic Press, New York, (doi: 10.1007/978-3-662-43208-2_2)

17) Boyko Y.B., Zabello E.I., Tikhonov E.A., 1980, *Ukr. J. of Physics*, **25,** 982-988

18) Zeidler G., **1980**, *J. Appl.Physics*, **42,** 884-885,